\documentclass [subeqn,twocolumn,showpacs,amsmath,prb]{revtex4}
\usepackage{amssymb}
\usepackage{epsfig}
\usepackage{graphicx}
\usepackage{natbib}

\begin{document}

\draft

\title{Phonon-induced electron relaxation in quantum rings}

\author{G. Piacente \cite{giovanni} and G.~Q. Hai \cite{hai}}

\address{Instituto de Física de S\~ao Carlos, Universidade de S\~ao Paulo,
13560-970 S\~ao Carlos, S\~ao Paulo, Brazil}

\date{}

\begin{abstract}

We study electron-acoustic phonon scattering and electron relaxation
in quantum rings in the absence and in the presence of external
magnetic fields. Electron-phonon interaction is accounted for both
the deformation potential and piezoelectric coupling. At zero
magnetic field and small ring radius, the deformation potential
phonon scattering is orders of magnitude larger than the
piezoelectric one. However, the piezoelectric coupling is found to
become important when the ring radius and magnetic field increase.
It can be the dominant scattering mechanism at large ring radii
and/or large magnetic fields. In comparison with the quantum dot
case, the acoustic phonon scattering is stronger in quantum rings.

\end{abstract}


\maketitle

\section{INTRODUCTION}

Recent successful fabrication of self-assembled quantum rings (QRs)
in nanometer dimensions \cite{garcia,lorke,mailly,mano} has
triggered a great deal of interest in theoretical and experimental
research. As a matter of fact, the QR structures could lead to the
development of novel devices in the fields of quantum cryptography,
quantum computation, optics and optoelectronics. These nanoscale
quantum objects also provide an excellent ground to investigate many
fundamental physical phenomena such as quantum coherence, which is
also critical for quantum information processing.

Study on electron energy relaxation in semiconductor quantum dots
(QDs) has received considerable attention
\cite{FujisawaNAT,FujisawaJPCM,FujisawaSCI,OrtnerPRB} due to the
critical importance of carrier-relaxation processes in the
performance of novel semiconductor devices (e.g., lasers and
infrared photo-detectors) based on QDs. One fundamental question is
how the discrete density of electronic states in these
nanostructures affects the relaxation time of excited electrons
\cite{BastardPRB,BockelmannPRB,moriPRB,stroscioPRB,kim1,kim2}. It
was recognized that electron-phonon scattering is severely
suppressed in QDs \cite{BastardPRB,BockelmannPRB}. In a QD where the
energy level separation is small as compared to the optical phonon
energy, the electron-acoustic phonon interaction is dominant.

Electron relaxation in QDs due to acoustic phonon scattering has
been studied extensively over the past few years
\cite{FujisawaNAT,FujisawaJPCM,FujisawaSCI,OrtnerPRB}. Coupling to
acoustic phonons has been effectively proved to be the most
important mechanism of electron relaxation from excited states in
laterally confined QDs \cite{FujisawaNAT,FujisawaJPCM} and
vertically coupled quantum dots (CQDs) \cite{FujisawaSCI,OrtnerPRB}.
Both the deformation potential (DF) and the piezoelectric (PZ)
acoustic phonons can scatter strongly the excited electrons. These
scattering mechanisms have been recently studied for single and
coupled QDs
\cite{ClimentePRB,StavrouPRB,WuPRB,BertoniAPL,BertoniPHE,HaiAPL}.
Theoretical estimations of phonon-induced scattering rates have
shown a satisfactory agreement with transport spectroscopy
experiments \cite{lorke}.

In this work, we extend previous calculations of the
electron-acoustic-phonon relaxation to the case of quantum rings. In
particular, we consider GaAs/AlGaAs QRs. We study single electron
relaxation in a QR as a necessary first step to further
investigations of the few-electron and the coupled quantum ring
cases. The electron is considered photo-excited or electrically
injected to an excited state. The ring is laterally confined and the
effects of the confinement and external magnetic fields are
investigated. We took into account both DF and PZ phonon scattering
mechanisms and we stress the differences between the QD and QR
geometries. The scattering rate demonstrates strong oscillations as
a function of the confinement and the external field.

The outline of our paper is as follows. In Sec. \ref{sec1} we
present our theoretical model and explain the numerical methods to
solve the single-particle problem. In Sec. \ref{sec2} we show the
numerical calculations of the relaxation rates in the absence and in
the presence of magnetic fields, separately. Based on our results,
we comment about the differences in the electron relaxation
processes between QRs and QDs. Finally, we conclude in Sec.
\ref{sec4}.

\section{Theoretical model}\label{sec1}

Within the effective mass and envelope function approximation we set
up the following model to study the electron relaxation time in QRs.
The confinement potential of the QR in the $xy$ plane is modeled by
a displaced parabolic function $V (\overrightarrow{\bf r}) =m^{\ast}
\omega_0^2 ( \overrightarrow{\bf r}_\parallel -  r_0)^2 / 2$, where
$m^{\ast}$ is the electron effective mass, $\overrightarrow{\bf
r}_\parallel \equiv(x,y)$, $r_0$ is the ring radius, and
$\omega_{0}$ describes the strength of the harmonic confinement in
the plane. The confinement potential in the $z$ direction $V_{QW}(z)
$ is considered as a quantum well of thickness $L_z$ and barrier
height $V_0$. When a magnetic field $\bf B $ is applied in the $z$
direction, the Hamiltonian for the single electron problem is given
by
\begin{equation}
\label{hamil_xy} H  = -\frac{(\overrightarrow{\mathbf p}+ e
\overrightarrow{\mathbf A})^2}{2m^{\ast}}
+\frac{1}{2}m^{\ast}\omega^{2}_{0}(\overrightarrow{\mathbf
r}_\parallel - r_0)^{2}+
V_{0}\Theta\left(\left|z\right|-L_{z}/2\right),
\end{equation}
where $\overrightarrow{\bf p} = -i \hbar {\overrightarrow{\bf
\nabla}}$, $\overrightarrow{\bf A} = (-By/2, Bx/2, 0)$, $\Theta(z)$
is the Heavyside function and $V_{0}$ the conduction band offset
between the GaAs well and the AlGaAs barrier.

Since the vertical confinement in the $z$ direction is usually much
stronger than the lateral one, the in-plane motion and the vertical
one can be treated as decoupled. The electron wavefunction can
thereby be written as a product:
\begin {equation}
\label{envelope} \psi(\overrightarrow{\bf{r}}) =
\psi_{\parallel}(\overrightarrow{\bf r}_{\parallel})
               \chi(z),
\end {equation}
where $\overrightarrow{\bf r}\equiv (x,y,z)$. The functions
$\psi_{\parallel}(\overrightarrow{\bf r}_{\parallel})$ and
$\chi(\it{z})$ describe the electron motion in the $xy$ plane and in
the $z$ direction, respectively.

It is known that the present eigenvalue problem of the QR is not
analytically solvable. Therefore, we evaluate the matrix elements of
the in-plane ring potential on the basis of the two-dimensional
parabolic QD, that is, on the Fock-Darwin basis, and diagonalize the
corresponding matrix \cite{simoninPRB}. Thus, the eigenvalues give
the energy levels of the in-plane motion and the eigenvector
components $\kappa_n^j$ yield the matrix of the change of basis. The
total electron eigenfunction is then a superposition of the
Fock-Darwin orbitals
\begin{equation}
\label{ring_fun} \psi_{j,m,g}(\overrightarrow{\bf r})= \Big( \sum_n
\kappa_n^j \phi_{n,m}(x,y) \Big) \chi_g(z)\; ,
\end{equation}
where $\phi_{n,m}$ is the $n$-th Fock-Darwin orbital with azimuthal
angular momentum $m$ and $\chi_g$ is the $g$-th finite quantum well
solution. The eigenstates can be labeled by the set of three quantum
numbers $(j,m,g)$, with $j=0,1,2,\ldots$, $m=0,\pm 1,\pm 2, \ldots$
and $g=0,1,2,\ldots$, respectively. Throughout this paper we assume
that the electron is ``frozen'' in the lowest energy state of the
quantum well along the $z$ direction, which means that we always
consider $g=0$. This assumption is justified by the fact that in
normal growth condition the vertical confinement is much stronger
than the in-plane one.

The electron-acoustic-phonon interaction Hamiltonian, disregarding
umklapp processes, is given by

\begin{equation}
\label{ep} H_{ep} = \sum_{\overrightarrow{\bf
q},\lambda}M_{\lambda}(\overrightarrow{\mathbf
 q})(a^{\dagger}_{\overrightarrow{\mathbf q}\lambda}+ a_{\overrightarrow{\mathbf
q}\lambda})\exp(i \overrightarrow{\mathbf q}\cdot
\overrightarrow{\mathbf r}),
\end{equation}

\noindent where $M_{\lambda}(\overrightarrow{\bf q})$ is the
scattering matrix element, $\overrightarrow{ \mathbf{q}}$ the phonon
wavevector, $a_{\overrightarrow{ \mathbf{q}} \lambda}$ and
$a_{\overrightarrow{\mathbf{q}}\lambda}^\dag$ the phonon
annihilation and creation operators, respectively, and $\lambda$ the
polarization index. At low temperature phonon absorption is
negligible. The electron scattering time between the initial state
$\psi^i(\overrightarrow {\bf{r}})$ and the final state
$\psi^f(\overrightarrow {\bf{r}})$ due to phonon emission can be
calculated according to the Fermi golden rule,
\begin{multline}
\label{eq1}
 \tau^{-1}_{if}=\frac{2\pi}{\hbar}\,\sum_{\overrightarrow{\bf q},\lambda}
  \lvert M_{\lambda}(\overrightarrow{\bf q})\rvert^2\, \lvert\langle
\psi^f|\,e^{-i \overrightarrow{\bf q}\cdot \overrightarrow{\bf
r}}\,|\psi^i \rangle\rvert^2\, \times \\ \delta(E_f - E_i -
\hbar\omega_q ),
\end{multline}

\noindent where $E_f$ and $E_i$ stand for the final and initial
electron levels, respectively, and $\hbar\omega_q$ is the phonon
energy. We study phonon modes with a linear dispersion.

In the considered GaAs semiconductor QR, an electron interacts with
the longitudinal acoustic (LA) phonon modes through a deformation
potential and with both the longitudinal and the transverse acoustic
(TA) phonon modes through piezoelectric fields
\cite{Mahan,Price,Mendez,Walukiewicz}. Thereby, the total scattering
matrix element is:

\begin{equation}
\label{Mtot} \lvert M (\overrightarrow{\bf q})\rvert^2 = \lvert
M_{\mbox{\tiny LA}}^{\mbox{\tiny DF}}(\overrightarrow{\bf
q})\rvert^2 + \sum_{\lambda {\mbox{\tiny =LA,TA}}} \lvert
M_{\lambda}^{\mbox{\tiny PZ}}(\overrightarrow{\bf q})\rvert^2,
\end{equation}

\noindent with the first contribution arising from the deformation
potential coupling and the second one from the piezoelectric
coupling. Notice that the Hamiltonian of the DF and PZ interactions
are real and imaginary, respectively, which allow us to investigate
these interactions separately and obtain the total contribution by
simply adding up the two rates \cite{Mahan}.

The electron-LA phonon scattering due to the deformation potential
is given by \cite{vogl,zhou}

\begin{equation}
\label{deform} \lvert M_{\mbox{\tiny LA}}^{\mbox{\tiny
DF}}(\overrightarrow{\mathbf{q}}) \rvert^2=\frac{\hbar D^2}{2\, \rho
\, c\, \Gamma}\,|\overrightarrow{\mathbf{q}}|,
\end{equation}

\noindent where $D$, $\rho$, $\Gamma$ and $c$ are the deformation
potential constant, the crystal density, the volume and, the
longitudinal sound velocity, respectively.

For zinc-blende crystal such as GaAs, the only non-vanishing
independent piezoelectric constant is $h_{14}$. The corresponding
scattering matrixes are given by \cite{vogl,zhou}

\begin{equation}
\lvert M_{\mbox{\tiny LA}}^{\mbox{\tiny PZ}}(\overrightarrow{\bf q})
\rvert^2=\frac{32 \pi^2\,\hbar\,e^2\,h_{14}^2}{\epsilon_0^2\,\rho
\,c \,\Gamma}\, \frac{(3\,q_x\,q_y\,q_z)^2}{|\overrightarrow{\bf
q}|^7}, \label{piezoLA}
\end{equation}
for the electron-LA-phonon scattering ($\lambda \equiv \,\,$LA), and
by

\begin{multline}
\label{piezoTA} \lvert M_{\mbox{\tiny TA}}^{\mbox{\tiny
PZ}}(\overrightarrow {\bf q }) \rvert^2=\frac{32 \pi^2\,
\hbar\,e^2\,h_{14}^2}{\epsilon_0^2\,\rho \,c'\,\Gamma} \, \times \\
\left| \frac{q_x^2\,q_y^2 + q_y^2\,q_z^2 +
q_z^2\,q_x^2}{|\overrightarrow {\bf q }|^5} -
\frac{(3\,q_x\,q_y\,q_z)^2}{|\overrightarrow {\bf q}|^7} \right|,
\end{multline}

\noindent for the electron-TA-phonon scattering ($\lambda \equiv
\,\,$TA), where $c'$ is the transversal sound velocity. In the above
Eqs.\ (\ref{deform}), (\ref{piezoLA}) and (\ref{piezoTA}) the phonon
dispersions are assumed as $\omega_{q,{\mbox{\tiny LA}}} =cq $ and
$\omega_{q,{\mbox{\tiny TA}}} =c'q$.

In order to obtain a more tractable form of the piezoelectric
coupling, we performed an angular average for the longitudinal and
transverse modes separately and then add the terms
\cite{Mahan,Zook}. This approximation leads to a minor quantitative
modification on the total scattering rate in comparison with
retaining the full $\overrightarrow {\bf q }$-dependence. It will
neither influence our main results nor the conclusions in this work.
We get the following expression for the
electron-piezoelectric-phonon scattering matrix elements:

\begin{equation}
\label{MpzAv} |M^{\mbox{\tiny PZ}}(|\overrightarrow {\bf q }|)|^2 =
\frac{\hbar \, e^2 \, h_{14}^2}{2\, \rho \, c\, \Gamma} \left(
\frac{12}{35} + \frac{1}{\alpha} \frac{16}{35}
\right)\frac{1}{|\overrightarrow {\bf q }|} \equiv \frac{\hbar}{2\,
\rho \, c\, \Gamma}\frac{P}{|\overrightarrow {\bf q }|},
\end{equation}

\noindent where we have introduced the constants $\alpha$ and $P$,
which represent the ratio between the longitudinal and transverse
sound velocities and the piezoelectric coupling constant,
respectively.

In this way, the total electron-phonon scattering matrix element in
Eq. (\ref{Mtot}) becomes,
\begin{equation}
 \lvert M (\overrightarrow
{\bf q })\rvert^2=\frac{\hbar}{2\, \rho \, c\,
 \Gamma} \frac{1}{|\overrightarrow
{\bf q }|} \big( D^2{|\overrightarrow {\bf q }|}^2+P\big).
\end{equation}
Consequently, the total scattering rate from $(j,m,g)$ to
$(j',m',g')$ state at $T=0$ can be written explicitly as,

\begin{multline}
\label{totalscatt} \tau^{-1}_{(j,m,g)\rightarrow (j',m',g')}=
\frac{D^2 q_0^3 +P q_0}{2 \pi \rho \hbar c^2} \sum_{n'= 0}^N
\sum_{n = 0}^N \kappa_{n'}^{j'} \kappa_n^j \, \, \times\\
\int_0^{\pi / 2}d\theta \sin \theta \lvert \langle
\phi_{n',m'}\lvert e^ {-i q_0 r
\sin \theta }\rvert \phi_{n,m}\rangle \rvert^2  \, \, \times \\
\lvert \langle \chi_g' \lvert e^ {-iq_0 z \cos \theta }\rvert \chi_g
\rangle \rvert^2,
\end{multline}

\noindent where $ q_0 = (E_{j',m',g'}- E_{j,m,g})/ \hbar c$ and $N$
is the number of the Fock-Darwin orbitals considered.

\begin{table}
\caption{The GaAs constants used in the paper. Unless otherwise
indicated the values are taken from Ref.~\onlinecite{blakemore}.}

\begin{ruledtabular}

\begin{tabular}{l@{\hspace{0mm}}c@{\hspace{0mm}}r@{\hspace{0mm}}l}
parameter & symbol & \multicolumn{2}{c}{value} \\ \hline effective
electron mass & $m^*$ & 0.067 & $\,$$m_0$\\crystal density &
$\rho$ & 5.3$\times$10$^{3}$ & $\,$kg m$^{-3}$ \\
band off-set &$V_0$ & $3.9\times$10$^{-20}$ & J\\
longitudinal sound velocity\cite{note} &
$c$ & 3.7$\times$10$^{3}$ & $\,$m s$^{-1}$ \\
transverse sound velocity\cite{note}&
$c'$ & 3.2$\times$10$^{3}$ & $\,$m s$^{-1}$ \\
static dielectric constant & $\epsilon_0$ & 12.8 &\\
deformation potential
& $D$ & 2.2$\times$10$^{-18}$ & $\,$J\\
piezoelectric constant&
$h_{14}$ & 1.38$\times$10$^{9}$ & $\,$V m$^{-1}$\\
piezoelectric coupling
& $P$ & 5.4$\times$10$^{-20}$  & $\,$J$^2$ m$^{-2}$ \\
\end{tabular}

\end{ruledtabular}

\label{GaAsConst}
\end{table}

\section{Numerical results and discussion}\label{sec2}

A quantum ring is distinctly different from a dot in that it has a
not-simply-connected geometry. This property makes QRs particularly
interesting, especially for studies in the presence of a vertical
magnetic field.

\begin{figure}[ht]
\begin{center}
\includegraphics [width=7.5cm]{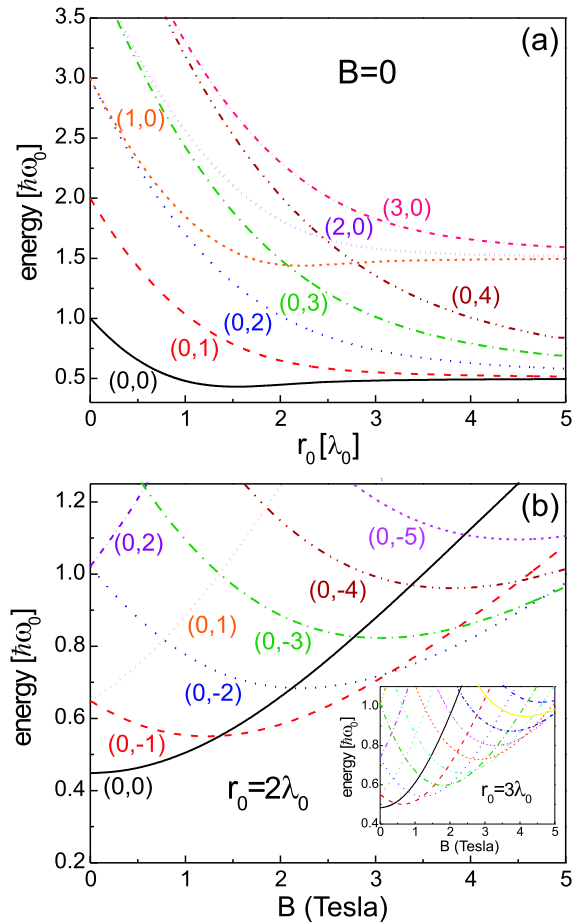}
\end{center}
\protect\caption{(Color online). The energy spectrum for the
electron in-plane motion as a function of (a) the ring radius at
$B=0$ and (b) the magnetic field for $r_0 = 2 \lambda_0$. The inset
in (b) shows the spectrum for a ring with $r_0 = 3\lambda_0$ in the
presence of magnetic field. For $B=0$ the electron levels have the
twofold degeneracy $\pm m$.} \label{fig1}
\end{figure}

First of all, we study the energy spectrum for the in-plane motion.
In Fig. \ref{fig1} we show the energy levels as a function of (a)
the ring radius at zero magnetic field and (b) the magnetic field at
fixed ring radii. The ring radius is measured in units of $\lambda_0
= \sqrt{\hbar/m^{\ast}\omega_{0}}$ and the energies in units of $
\hbar \omega_{0} $. For a lateral confinement $\hbar\omega_{0}=5$
meV, one gets $\lambda_0 \sim 15$ nm, which is a typical value for
the ring radius reported in recent experiments \cite{emperador}. In
Fig. \ref{fig1}(a) we reproduced the results already obtained in
Ref.~\onlinecite{simoninPRB} within an identical theoretical
framework. We got results in perfect agreement with the
aforementioned ones, as one can see immediately comparing Fig.
\ref{fig1}(a) to Fig. 1 in Ref.~\onlinecite{simoninPRB} in the case
of zero magnetic field. Moreover, in the presence of magnetic field,
we calculated the spectra for values of the ring radius other than
the ones shown in Fig. 4 of Ref.~\onlinecite{simoninPRB}.

It is interesting to notice the differences between the cases
$B$$=\,$0 and $B$$\neq \,$0. In the absence of magnetic field the
lowest states have definite quantum numbers (although the degeneracy
$\pm m$ is present). When a vertical magnetic field is present,
level crossings occur due to the multiple connected geometry of QRs.
The ground state changes from the state with $m$$=\,$0 to the ones
with $m=-1,-2,-3,...$ with increasing magnetic field (see Fig.
\ref{fig1}(b)). The ring eigenfunctions are piecewise defined as a
function of $B$. This feature is not present in a QD where no level
crossings occur between the ground state and the low-lying levels.
They converge smoothly to the first Landau levels at large magnetic
fields. Actually, identifying the crossing points in the energy
spectra in a QR gives straightforward information about the specific
ring topology \cite{climenteR,fuhrer}. Notice that the crossing
point density in magnetic field increases as the ring radius
increases.

In what follows, we report the calculations of the electron
relaxation time due to acoustic phonon scattering in the cases
$B$$=\,$0 and $B$$\neq\,$0. We consider the relaxation rates from
the first and second excited states to the ground state because they
are often the most relevant transitions. These transitions can be
monitored experimentally (for instance, by means of pump-and-probe
techniques \cite{FujisawaNAT,FujisawaJPCM}). For an electron
initially in the first excited state, the relaxation rate is given
by the direct scattering rate from this state to the ground state.
For electrons in the second (or higher) excited state, different
relaxation channels exist and the relaxation rate can be obtained
using Matthiessen's rule.

\subsection{In the absence of magnetic field}\label{sec3a}

In this section we study the acoustic phonon scattering and electron
relaxation rate in QRs at zero magnetic field. Electron-acoustic
phonon interactions in nano-structures are in general determined by
the interplay of lateral confinement length, quantum well thickness
and phonon wavelength \cite{BockelmannPRB}.

\begin{figure}[htpb]
\begin{center}
\includegraphics [width=7.3cm]{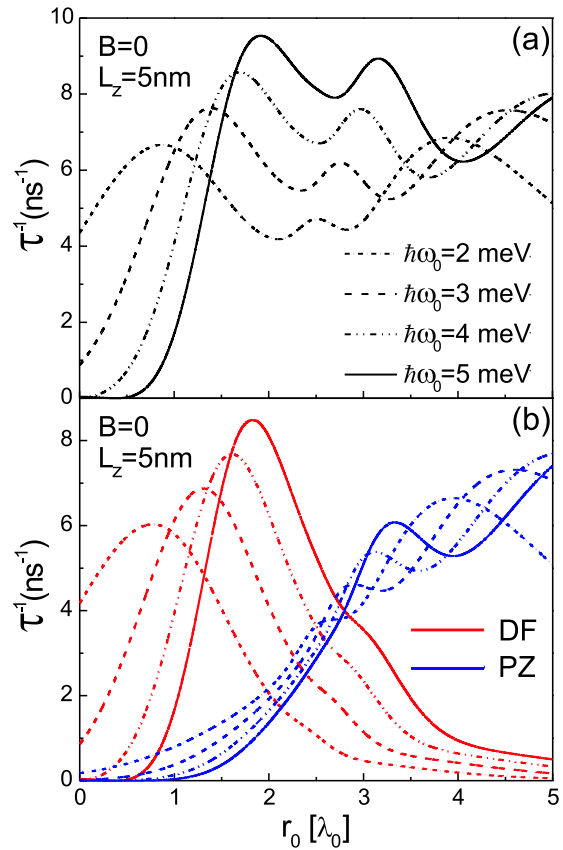}
\end{center}
\protect\caption{(Color online). (a) The total scattering rate from
$(0,\pm 1,0)$ to $(0,0,0)$ state as a function of the ring radius
for different lateral confinements $\hbar \omega_0$. The vertical
well thickness is $L_z=5$ nm. (b) The contributions of the DF and PZ
couplings are plotted separately.} \label{fig2}
\end{figure}

First of all, we calculate the scattering rate as a function of the
ring radius $r_0$ for different lateral confinement frequencies. The
scattering rate from the first excited state $(0,\pm1,0)$ to the
ground state $(0,0,0)$ is shown in Fig.~\ref{fig2}(a) for a fixed
ring thickness $L_z=5$ nm. The scattering rate $\tau^{-1}$ shows an
oscillatory behavior as a function of the ring radius. For very
small $r_0$ (corresponding to quantum dots), the total scattering
rate is small and decreases as the confinement frequency $\omega_0$
increases. With increasing the ring radius, the scattering rate
increases pronouncedly for small $\omega_0$. As is evident in
Fig.~\ref{fig2}(b), for the considered confinement frequencies, the
DF acoustic phonon scattering is much larger than the PZ phonon
scattering for small ring radius. This is a well known result for
QDs where the DF phonon scattering is dominant \cite{BockelmannPRB}.
However, our calculation shows that the PZ phonon scattering becomes
much larger than that of the DF in QRs (large $r_0$). For each
$\hbar \omega_0$, there is a clear crossover point from which the PZ
scattering rate starts to dominate over the DF one. Therefore, in
QRs, the PZ effects are important and cannot be neglected as it is
usually done in QDs. As one can see from Eq. (\ref{totalscatt}), the
relative contributions of DF and PZ scattering are determined by the
prefactor before the integral. At small $q_0$ the PZ coupling is
dominant over the DF one. Actually, by increasing the ring radius
the energy levels get closer and closer (see Fig. \ref{fig1}(a)),
and consequently $q_0$ becomes smaller and smaller. This is the
reason why after a certain $r_0$ the PZ coupling becomes the
dominant scattering mechanism. We will return and comment more on
this point in the next section.

Although the PZ and DF scattering rates are of different features,
their behavior in the QD limit (i.e., $r_0 \rightarrow 0$) is
similar. They decreases rapidly as the lateral confinement frequency
increases. The reason is that the orthogonality of the electron
states leads the factor $ \langle \psi^f \lvert \, \exp({-i
\overrightarrow{\mathbf q} \cdot \overrightarrow{\mathbf r} })\lvert
\psi^i \rangle$ to vanish rapidly when the phonon wavelength is
shorter than the lateral confinement length \cite{BockelmannPRB}.

\begin{figure}[ht]
\begin{center}
\includegraphics [width=7.5cm]{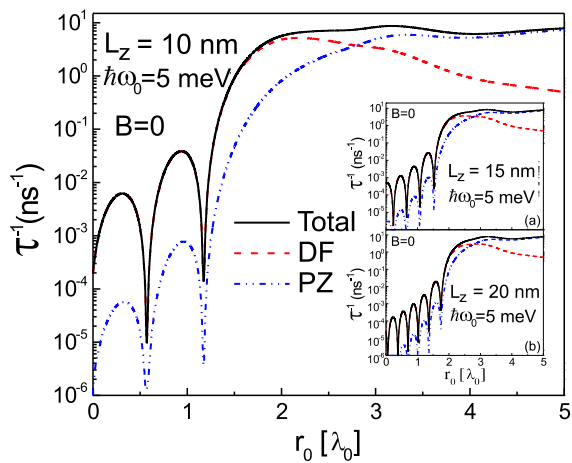}
\end{center}
\protect\caption{(Color online). The total, DF and PZ scattering
rates from $(0,\pm 1,0)$ to $(0,0,0)$ state as a function of the
ring radius for different vertical confinements $L_z$. The lateral
confinement is $\hbar \omega_0 =5$ meV.} \label{fig3}
\end{figure}

In Fig. \ref{fig3}, the effect of the QR thickness $L_z$ on the
phonon scattering rates is studied. The DF, PZ, as well as total
scattering rates are given as a function of the ring radius for
three different $L_z$ at fixed lateral confinement $\hbar
\omega_0=\,$5nm. The scattering curves exhibit strong oscillations
for small ring radius. The oscillation in the electron-phonon
scattering rate is originated by the match of the localized electron
wavelength to the wavelengthes of the emitted phonons in the
scattering as discussed in Ref.~\onlinecite{BockelmannPRB} for the
QD case. In the calculation, the electron wavefunction is decoupled
in the $z$ direction and in the $xy$ plane. Thus, such a match
depends on the projection of the phonon wavevector onto the $xy$
plane due to its 3D characteristic. For large $L_z$, the
electron-phonon scattering elements depend more strongly on the
projection angle of the phonon wavevector onto the $xy$ plane.
Consequently, the oscillation in the scattering rate is stronger for
larger $L_z$. It reaches maxima (minima) when the electron wave
function is in-phase (anti-phase) with the phonon wave. The
scattering rate can be suppressed by orders of magnitude due to the
anti-phase relation between the electron wave function and the
phonon wave in the ring. This effect has been proposed as a possible
way to control the coherence time for single and coupled QDs
\cite{ClimentePRB,BertoniAPL,BertoniPHE}. With increasing the ring
radius, the phonon scattering is enhanced pronouncedly and the
oscillations disappear.

\begin{figure}[ht]
\begin{center}
\includegraphics [width=7.5cm]{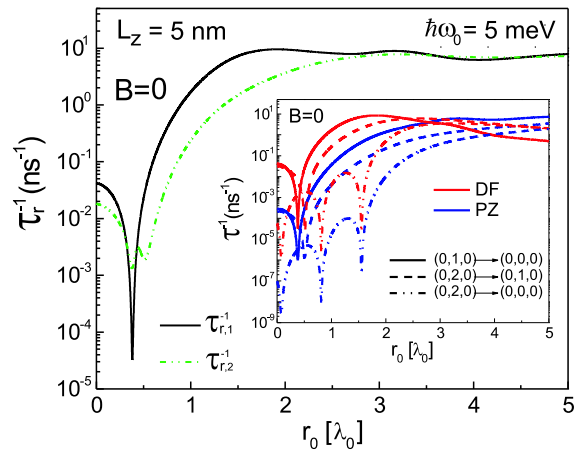}
\end{center}
\protect\caption{(Color online). The relaxation rates for the
transitions from the first excited state and the second excited
state to the ground state as a function of the ring radius. The
inset shows the DF and PZ scattering time for all the transitions
involved in the relaxation processes.} \label{fig4}
\end{figure}

We calculated the relaxation rates $\tau_r^{-1}$ for the transitions
from the first and second excited states to the ground state as
shown in Fig. \ref{fig4} for the QRs with $\hbar \omega_0=5$ meV and
$L_z=5$ nm. For an electron in the first excited state $(0,\pm1,0)$,
its relaxation rate $\tau_{r,1}^{-1}$ to the ground state coincides
with the direct scattering rate between the two states, as already
mentioned. For an electron in a higher state, its relaxation to the
ground state is in general a multichannel process. In the specific
case under investigation, an electron in the second excited state
$(0,\pm2,0)$ can relax to the ground state by direct transition
$(0,\pm2,0)\rightarrow(0,0,0)$ or indirect transition
$(0,\pm2,0)\rightarrow(0,\pm1,0)\rightarrow(0,0,0)$. Using
Matthiessen's rule, the relaxation rate for the second excited state
can be written as

\begin{equation}
\label{matth}
\begin{split}
 & \tau_{r,2}^{-1}= \tau_{(0,\pm2,0)\rightarrow(0,0,0)}^{-1} \,\,\,+   \\&
(\tau_{(0,\pm2,0)\rightarrow(0,\pm1,0)}+\tau_{(0,\pm1,0)\rightarrow(0,0,0)})^{-1}.
\end{split}
\end{equation}

From Fig. \ref{fig4} one can notice that for large ring radius the
two relaxation rates $\tau_{r,1}^{-1}$ and $\tau_{r,2}^{-1}$ are
very similar and of the order of 10 ns$^{-1}$. For small ring
radius, interesting features appear. The relaxation rate
$\tau_{r,1}^{-1}$ shows a striking dip at $r_0\sim0.38\lambda_0$ and
$\tau_{r,2}^{-1}$ has two minima at $r_0\sim0.38\lambda_0$ and
$r_0\sim0.52\lambda_0$ where the relaxation rate is reduced by
orders of magnitude. This effect results from the anti-phase
relation between the electron and phonon wave functions as discussed
above. Although it is extremely difficult to control QR geometry in
the fabrication process, in the same spirit of
Refs.~\onlinecite{ClimentePRB,BertoniAPL,BertoniPHE} we can affirm
that a geometry-induced control of the relaxation time is feasible
also in the case of QRs of small radius.

The inset of Fig. \ref{fig4} provides us with some extra
information: the DF coupling is more efficient than the PZ one in
the relaxation processes from higher excited states with respect to
the first excited state. In other words, the crossing where the PZ
scattering starts to dominate over the DF one in the transitions
$(0,\pm2,0)\rightarrow(0,0,0)$ and $(0,\pm2,0)\rightarrow(0,\pm1,0)$
takes place at a larger $r_0$ than that in the transition
$(0,\pm1,0)\rightarrow(0,0,0)$.

\begin{figure}[ht]
\begin{center}
\includegraphics [width=7.5cm]{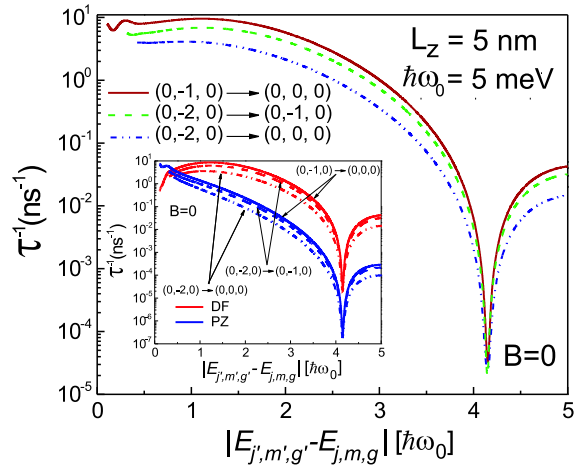}
\end{center}
\protect \caption{(Color online). The scattering rate for the
transitions between two low-lying adjacent states as a function of
the energy difference of the two levels. In the inset the DF and PZ
contributions for all the possible transitions are given
separately.} \label{fig5}
\end{figure}

To conclude this subsection, we plot in Fig. \ref{fig5} the
scattering rates for the transitions $(0,\pm1,0)\rightarrow(0,0,0)$,
$(0,\pm2,0)\rightarrow(0,\pm1,0)$ and $(0,\pm2,0)\rightarrow(0,0,0)$
as a function of the energy difference between the corresponding
levels (i.e., the emitted phonon wavenumber
$q_0=(E_{j',m',g'}-E_{j,m,g})/\hbar c$)instead of the ring radius.
Though the scattering rate is strongly dependent on the initial and
final states, it is evident from Fig. \ref{fig5} that the scattering
rates for the same $q_0$ are very similar. This indicates that a
crucial quantity in the scattering process is the energy difference
between the ring levels or $q_0$. Furthermore, strong scattering
occurs at small $q_0$ (i.e., long-wavelength phonons) where the
relaxation time is in the order of 0.1 ns. As one can see from the
inset, the PZ and DF phonon scattering in this region is of
approximatively the same contribution. This again confirms that
differently from the QD case, the PZ phonon scattering is critical
in the QRs. The scattering rate decreases rapidly with increasing
$q_0$ and the DF phonon scattering becomes dominant over the PZ one
(about two orders of magnitude). Strong scattering at small energies
implies that the transition probability for an electron from one
state to its adjacent level in energy is much larger than to other
levels. As a consequence, relaxation from a higher excited state to
the ground state is basically a multi-scattering process through all
the intermediate states between them.

\subsection{In the presence of magnetic field}\label{sec3b}

In this section we study the effects of a magnetic field $B$ on the
electron-phonon scattering and the electron relaxation time in a QR.
As shown in Fig. \ref{fig1}(b), an external magnetic field in the
$z$-direction alters drastically the electron energy spectrum.

For instance, the ground state in the case $r_0=2\lambda_0$ is
defined by the orbitals: $(0,0,0)$ for $0\leqslant B < 1.35$,
$(0,-1,0)$ for $1.35\leqslant B < 3.14$, $(0,-2,0)$ for
$3.14\leqslant B < 4.77$ and so on. Hence, the ring eigenstates as a
function of magnetic field are piecewise. In
Ref.~\onlinecite{gudmundsson} the effect of a time varying magnetic
field are shown to produce non-adiabatic changes upon the electron
states of the QR, leading to strong non-linear effects in the
time-depending magnetization.

\begin{figure}[ht]
\begin{center}
\includegraphics [width=7.3cm]{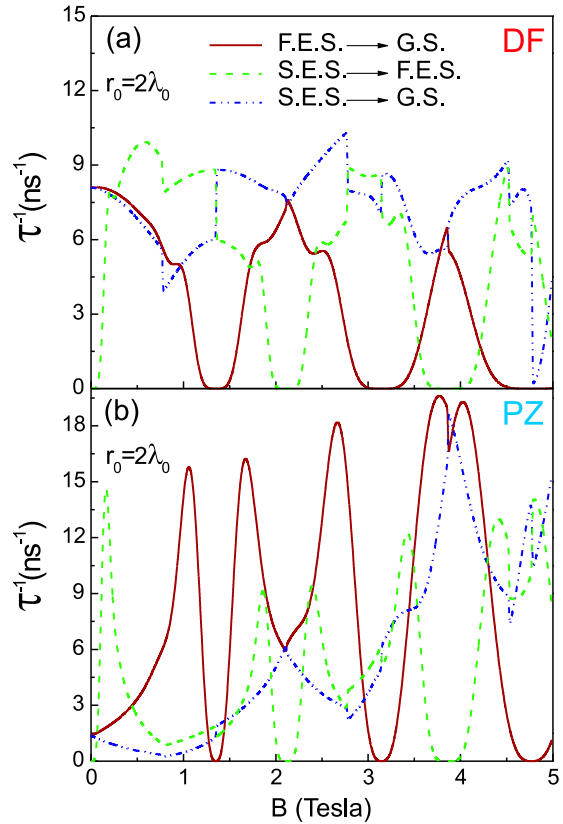}
\end{center}
\protect\caption{(Color online). The scattering rates as a function
of the magnetic field for a ring of radius $r_0=2\lambda_0$ for (a)
the DF and (b) the PZ couplings.} \label{fig6}
\end{figure}

\begin{figure}[ht]
\begin{center}
\includegraphics [width=7.3cm]{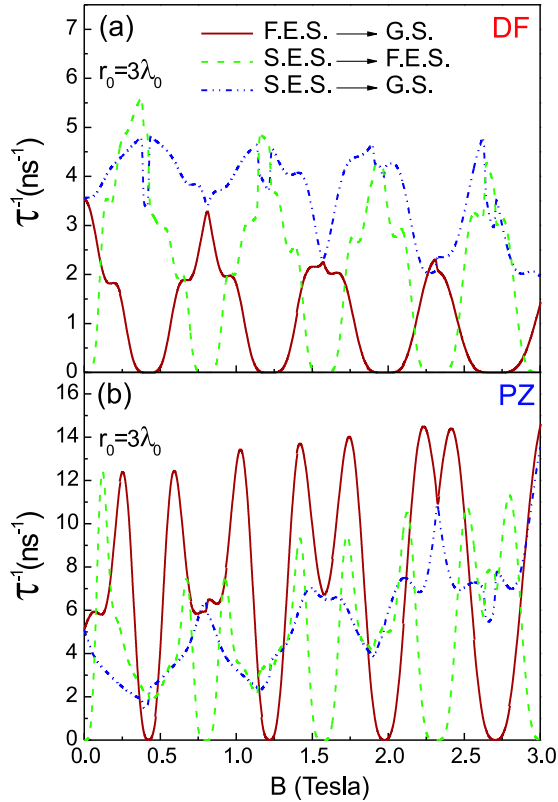}
\end{center}
\protect\caption{(Color online). The same as Fig. \ref{fig6}, but
with $r_0=3\lambda_0$.} \label{fig7}
\end{figure}

In what follows, we will refer to the ground state, the first and
second excited states as G.S., F.E.S. and S.E.S, respectively,
rather than as a set of three quantum numbers. From now on we
concentrate on a well defined ring geometry with $L_z$$=\,$5 nm and
$\hbar \omega_0$$=\,$5 meV. For fixed ring radii $r_0=2\lambda_0$
and $r_0=3\lambda_0$, we calculated the scattering rates among the
lowest three states (i.e., the S.E.S., F.E.S. and G.S.) as a
function of magnetic field.  The results are shown in Fig.
\ref{fig6} and Fig. \ref{fig7}, respectively. The contributions of
the DF and PZ phonons to the scattering rates are given separately.

The first notable feature is that the scattering rates due to the PZ
and DF phonons have approximatively the same order of magnitude in
magnetic fields. Therefore, for QRs in the presence of $B$ we cannot
disregard any one of the two scattering mechanisms. In order to
describe properly the phonon scattering processes in the QRs in
magnetic fields, inclusion of the PZ interaction is fundamental. In
fact, as is evident in Fig. \ref{fig6} and Fig. \ref{fig7}, the PZ
scattering is almost twice larger than the DF one for the transition
F.E.S. $\rightarrow$ G.S. Actually, the presence of a magnetic field
produces a similar effect as increasing the ring radius, that is it
draws the energy levels up, reducing in this way the value of $q_0$
and, consequently, enhancing the relative importance of the PZ
coupling.

The highest scattering rates in a small
ring ($r_0=2\lambda_0$) are larger than those in a larger ring
($r_0=3\lambda_0$). Furthermore, the scattering rates oscillate more
as a function of $B$ in the ring of $r_0=3\lambda_0$. This is
strictly related to the increasing number of level crossings as
$r_0$ becomes larger. Notice that in this case the oscillations are
due to the in-plane level crossings at which $q_0 \to 0$ rather than
to the anti-phase relation between the electron and phonon
wavefunctions in the QD case \cite{ClimentePRB}.

\begin{figure}[ht]
\begin{center}
\includegraphics [width=7.5cm]{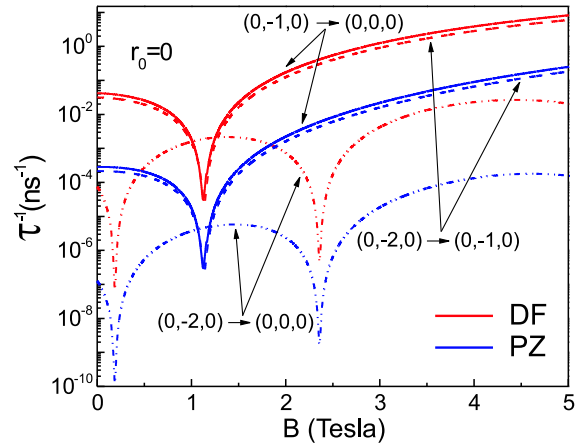}
\end{center}
\protect\caption{(Color online). The scattering rates as a function
of magnetic field in the case of a quantum dot.} \label{fig8}
\end{figure}

It is interesting to make a direct comparison with the QD case. In
Fig. \ref{fig8} the different scattering contributions are
calculated for a QD with $\hbar\omega_0=5$ meV and $L_z=5$ nm in
vertical magnetic fields, \emph{i.e.} the same thickness and
confinement frequency as the ring studied above. For magnetic fields
up to 5 Tesla, the PZ phonon scattering rates in the QD are always
orders of magnitude smaller than the DF ones, as one can see in Fig.
\ref{fig8}. Once more, this confirms that the phonon scattering
mechanism in QRs is significantly different the QDs. Such a
difference is amplified in the presence of magnetic fields. In the
QDs, the PZ phonon scattering can be neglected in the presence of
$B$. In the QRs, on the contrary, the PZ interactions are as important as
the DF ones.

\begin{figure}[ht]
\begin{center}
\includegraphics [width=7.5cm]{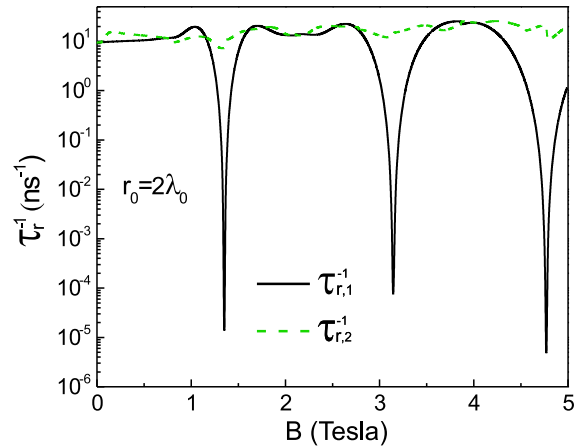}
\end{center}
\protect\caption{(Color online). The relaxation rate for the
transitions from F.E.S to G.S. and the S.E.S. to G.S. as a function
of magnetic field for a ring of radius $r_0=2\lambda_0$.}
\label{fig9}
\end{figure}

\begin{figure}[ht]
\begin{center}
\includegraphics [width=7.5cm]{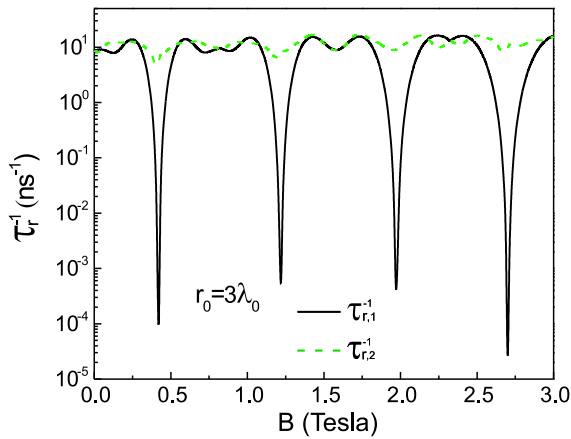}
\end{center}
\protect\caption{(Color online). The same as Fig. \ref{fig9} but for
$r_0=3\lambda_0$.} \label{fig10}
\end{figure}

In Fig. \ref{fig9} and Fig. \ref{fig10} we show the relaxation rates
from the F.E.S. to G.S. and from the S.E.S. to G.S. as a function of
$B$ for $r_0=2\lambda_0$ and $r_0=3\lambda_0$, respectively. In both
cases the relaxation rate from the F.E.S. to G.S. shows dips at the
level crossings between the two states, where the phonons (with
$q_0\to 0$) involved in the scattering are of vanishing density of
states. This mechanism is distinctly different from the suppression
of the scattering rate due to the anti-phase relation between the
electron and the phonon wave functions as was shown in Fig.
\ref{fig4}.

Variation of the relaxation rate from the second excited state to
the ground state in magnetic fields is quite small and no dips
appear. The reason is that $\tau^{-1}_{r,2}$ results from a
multi-scattering process and it is calculated according to
Matthiessen's rule in Eq. (\ref{matth}). Due to the fact that the
crossing points in the energy levels for the G.S., F.E.S. and S.E.S.
take place at different magnetic fields, the direct scattering rates
$\tau_{{\mbox{\tiny F.E.S.}}{\tiny\rightarrow}{\mbox{\tiny G.S.}}
}^{-1}$, $\tau_{ {\mbox{\tiny S.E.S.}} {\tiny \rightarrow}
{\mbox{\tiny F.E.S.}}}^{-1}$ and $\tau_{ {\mbox{\tiny S.E.S.}}
{\tiny \rightarrow} {\mbox{\tiny G.S.}}}^{-1} $ are never zero
simultaneously.

\begin{figure}
\begin{center}
\includegraphics [width=7.5cm]{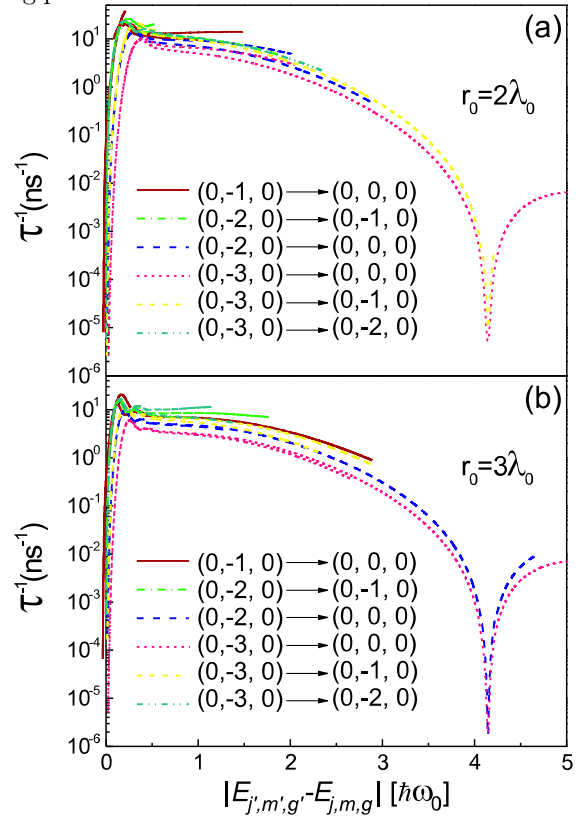}
\end{center}
\protect\caption{(Color online). The scattering rates in the
presence of a vertical magnetic field for the transitions between
the different levels defining the low-lying ring states as a
function of the energy differences. In (a) is depicted the case
$r_0=2\lambda_0$ and in (b) the case $r_0=3\lambda_0$. Notice the
similarity with Fig. \ref{fig5}.}\label{fig11}
\end{figure}

Finally, in Fig. \ref{fig11} we show the total scattering rates in
the presence of $B$ between two different levels as a function of
the energy difference between the corresponding levels instead of
magnetic field. Both the DF and PZ phonons are included. We observe
that the scattering rates in magnetic fields depends strongly on the
energy difference between the levels considered. But, at the same
energy difference, the different scattering rates are in the same
order of magnitude independently on the initial and final states.
Furthermore, when we compare Figs. \ref{fig11}(a) and \ref{fig11}(b)
of different ring radii (see also Fig. 5), we find that the curves
in the different figures are very similar to each other, i.e., the
values of the scattering rates and the positions of the maximum and
minimum are almost the same. We can, thereby, confirm that the
energy difference between the two states involved in the transitions
is a crucial parameter in determining the acoustic phonon scattering
rates. Changing of the ring radius and/or the magnetic field affects
essentially the energy separation between different levels and,
consequently, the phonon scattering rate.

\section{Conclusions}\label{sec4}

We studied the acoustic-phonon-induced relaxation of an excited
electron in a single GaAs quantum ring. We investigated the effects
of the ring geometry and external magnetic fields on the relaxation
from the first and second excited states to the ground state. Our
calculations show that electron-phonon scattering strongly depends
on the ring lateral and vertical dimensions and on the external
fields. We took into account both the deformation potential and
piezoelectric field couplings and shown that they both give
important contributions to the electron relaxation. The
piezoelectric interaction, which is often negligible in quantum
dots, represent the major source of scattering in quantum rings of
large ring radius or when a high magnetic field is applied.
For small ring radii, the deformation potential coupling prevail,
being similar to the quantum dot case.
Nevertheless, we have shown that the electron-phonon scattering rate
between any two levels are mainly determined by the energy
difference between them. In magnetic fields, a multiple scattering
process leads to the relaxation time in the order of $10^{-1}$
nano-second for an electron from the second excited state to the
ground state in QRs.

\section{Acknowledgments}

This work was supported by FAPESP and CNPq, Brazil.

\end{document}